\documentclass[11pt]{article}
\usepackage{palatcm, color, verbatim}

\usepackage{amssymb}
\usepackage{amsmath}
\usepackage{epsfig}
\usepackage[usenames,dvipsnames]{xcolor}
\usepackage{mathdots}

\begin{document}
\newtheorem{theorem}{Theorem}
\newtheorem{corollary}{Corollary}
\newtheorem{conjecture}{Conjecture}
\newtheorem{definition}{Definition}
\newtheorem{lemma}{Lemma}
\newtheorem{remark}{Remark}
\newcommand{\define}{\stackrel{\triangle}{=}}
\newcommand{\xS}{\mathbf{S}}
\newtheorem{construction}{Construction}
\newtheorem{property}{Property}

\newcommand\blfootnote[1]{%
  \begingroup
  \renewcommand\thefootnote{}\footnote{#1}%
  \addtocounter{footnote}{-1}%
  \endgroup
}

\pagestyle{empty}

\def\QED{\mbox{\rule[0pt]{1.5ex}{1.5ex}}}
\def\proof{\noindent{\it Proof: }}

\date{}

\title{Optimality of Orthogonal Access \\for One-dimensional Convex Cellular Networks}

\author{Hamed Maleki, Syed A. Jafar\\
%{\small hmaleki@uci.edu, viveck@mit.edu, syed@uci.edu}
\blfootnote{Hamed Maleki and Syed Jafar (email: hmaleki@uci.edu, syed@uci.edu) are with the Center for Pervasive Communications and Computing (CPCC) at the University of California Irvine, Irvine, CA, 92697.}}

%\IEEEauthorblockA{School of Electrical and\\Computer Engineering\\
%Georgia Institute of Technology\\
%Atlanta, Georgia 30332--0250\\
%Email: http://www.michaelshell.org/contact.html}
%\and
%\IEEEauthorblockN{Viveck Cadambe}
%%\IEEEauthorblockA{Twentieth Century Fox\\
%\and
%\IEEEauthorblockN{Syed A. Jafar}
%\IEEEauthorblockA{Starfleet Academy\\
%San Francisco, California 96678-2391\\

\maketitle
\thispagestyle{empty}
\begin{abstract}
It is shown that a greedy orthogonal access scheme achieves the sum degrees of freedom of all one-dimensional  (all nodes placed along a straight line) convex cellular networks (where cells are convex regions) when no channel knowledge is available at the transmitters except the knowledge of the network topology.  In general, optimality of orthogonal access holds neither for two-dimensional convex cellular networks nor for one-dimensional non-convex cellular networks, thus revealing a fundamental limitation that exists only when both one-dimensional and convex properties are simultaneously enforced, as is common in canonical information theoretic models for studying cellular networks. The result also establishes the capacity of the corresponding class of index coding problems.

%We consider a class of index coding problems inspired by linear cellular systems. We prove that a greedy solution is optimal to achieve the sum capacity of this class when we have two convex properties called transmitter and receiver convex properties.
%We consider a linear cellular system where there are arbitrary number of users and arbitrary number of cells on a line. The base stations have no knowledge of channel gains but they know the connectivity. We prove that a greedy orthogonal solution is optimal to achieve the sum-DoF of this network when we have two convex properties called message convex property and connectivity convex property.
\end{abstract}

\newpage

%\IEEEpeerreviewmaketitle

\section{Introduction}
\begin{figure}[!h] \centering
\includegraphics[width=3in]{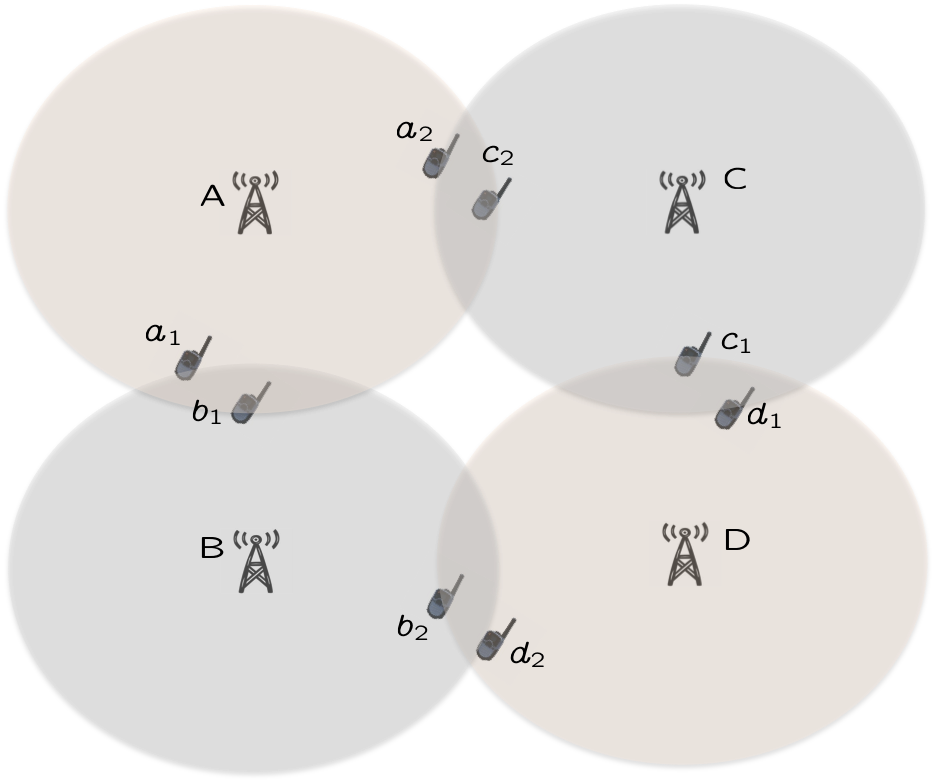}
\caption{\it A 4-cell example with no CSIT except topology where interference alignment is needed. Each base station ($A, B, C, D$) can only be heard within the shaded region around it.\cite{Jafar_TIM}}\label{fig:4cell}
\end{figure}

Interference is one of the main barriers faced by cellular networks. While many sophisticated interference management schemes have recently been developed around the idea of interference alignment, typically these schemes assume  too much channel state information at the transmitters (CSIT) to be considered practical. A complementary perspective, called topological interference management (TIM), is introduced in \cite{Jafar_TIM}, which studies the degrees of freedom (DoF) of partially connected wireless networks with only a knowledge of the network topology available to the transmitters. Essentially, TIM generalizes the idea of interference avoidance, by using optimal vector space assignments instead of  conventional solutions such as TDMA, FDMA or CDMA. The optimal assignment of vector spaces based only on the network topology, is shown in \cite{Jafar_TIM} to be essentially the ``index coding" problem that has been studied by computer scientists for over a decade. Optimal solutions to the index coding problem, and therefore  optimal assignments of vector spaces in  partially connected wireless networks, are shown to be  guided by interference alignment principles, even though no channel knowledge other than the connectivity is available to the transmitters in the wireless setting. Conventional access approaches such as TDMA/FDMA (which correspond to fractional clique covers in the index coding problem) and CDMA (which corresponds to  partition multicast) are special cases of the solutions proposed for index coding, and are sub-optimal in general. Reference \cite{Jafar_TIM} points out  instances of wireless interference networks with no CSIT except topology, where the optimal solution beats the  best TDMA/FDMA/CDMA solution by at least a factor of $(1+o(1))K^{1/4}$, where $K$ is the number of messages. Even for small cellular networks, such as the 4 cell example of \cite{Jafar_TIM} illustrated in Fig. \ref{fig:4cell}, the benefits of interference alignment over TDMA/FDMA/CDMA are evident. 
What is shown in Fig. \ref{fig:4cell} is a 4 cell network with 8 users, each located at the boundary of two cells. Users are labeled by the symbols that they desire. It is easy to see that the best TDMA/FDMA/CDMA scheme cannot achieve more than a total of 2 DoF in this network (e.g., by allowing base stations $A$ and $D$ to transmit while base stations $B, C$ remain silent). However, a simple interference alignment scheme with no CSIT  beyond network topology, i.e., the same CSIT as needed for TDMA, achieves $\frac{8}{3}$ DoF which is also the optimal DoF value. The optimal scheme operates over 3 channel uses over which base station A transmits $[a_1, a_2,0]$, B transmits $[b_1, b_1+b_2, b_1]$, C transmits $[c_1, 0, c_2]$, and D transmits $[d_1, d_1, d_1+d_2]$. Because of interference alignment, each user is able to resolve its desired symbol.

While TIM and index coding problems are open in general, for several interesting topologies the optimal solutions have been found. These include symmetric cellular settings with neighboring interferers or neighboring antidotes \cite{Index_coding_Maleki_Cadambe_Jafar}, symmetric linear, rectangular and hexagonal cellular network topologies \cite{Jafar_TIM}, settings with 5 or fewer messages \cite{Arbabjolfaei}, a sub-class of 6-cell topologies \cite{Naderializadeh_Avestimehr}, all half-rate-feasible networks \cite{Blasiak_Kleinberg_Lubetzky_2010, Jafar_TIM}, networks where the sum-DoF is only 1 \cite{Birk_Kol, Neely, Jafar_TIM}, and  settings with only one-to-one alignment demands where the alignment graph contains no overlapping cycles \cite{Sun_Jafar}. Even when the optimal solutions are orthogonal, e.g,. TDMA schemes, they are guided by interference alignment principles. For example, in the regular hexagonal cellular topology where each user is at the boundary of two cells whose base stations are the only ones it can hear, the optimal solution is an orthogonal scheme that achieves the optimal DoF value $\frac{6}{7}$ by sacrificing one cell towards which all interference is directed from its six neighboring cells, in effect aligning interference in the direction of the sacrificed cell. %Orthogonal solutions  possess a desirable robustness property  because they do not rely on channel coherence, i.e., they work even if the channel changes with every channel use. 

Finding solutions to new classes of topologies for the index coding and TIM problems is currently an active research area \cite{Blasiak_Kleinberg_Lubetzky_2010, Index_coding_Maleki_Cadambe_Jafar, Jafar_TIM, Arbabjolfaei, Naderializadeh_Avestimehr},. Evidently,  topologies inspired from cellular wireless networks are of particular interest. In this work, we  settle the TIM and index coding problems for such a class of topologies.

\section{Main Result}
We settle the DoF of all one-dimensional convex cellular networks when no CSIT is available except network topology. We define the terminology as follows.

\subsection{One-dimensional Convex Cellular Networks}

\noindent{\bf One-dimensional Network:}  A one-dimensional network  corresponds to a placement of all transmitters and receivers along a straight line. 

%Convex connectivity is defined as follows.
%By itself, a one-dimensional model carries no loss of generality, since any topology can be mapped to a one-dimensional topology by numbering the nodes and placing them sequentially. The class of topologies that we consider arises when the connectivity is consistent with a \emph{physical} placement of nodes along a straight line, specifically in the sense of convex connectivity. 

\bigskip

\noindent{\bf Convex Cellular Network Model:} Convex cellular network model captures propagation path loss and cell association properties, stated below from the destination's perspective.
\begin{enumerate}
\item {\it Propagation path loss:} Around each destination node $D_j$ there is a convex region that contains all the source nodes that $D_j$ can hear and none of the source nodes that $D_j$ cannot hear. 
\item {\it Cell Association:}  Around each destination node there is a convex region that contains all source nodes associated with $D_i$ (i.e., all source nodes that have desired messages for $D_j$) and none of the source nodes that are not associated with $D_j$.  
\end{enumerate}
A convex connected network should  satisfy corresponding properties from the source nodes' perspective as well. Note that there is no assumption of symmetry, so each convex region may be different.

\begin{figure}[!h] \centering
\includegraphics[width=6in]{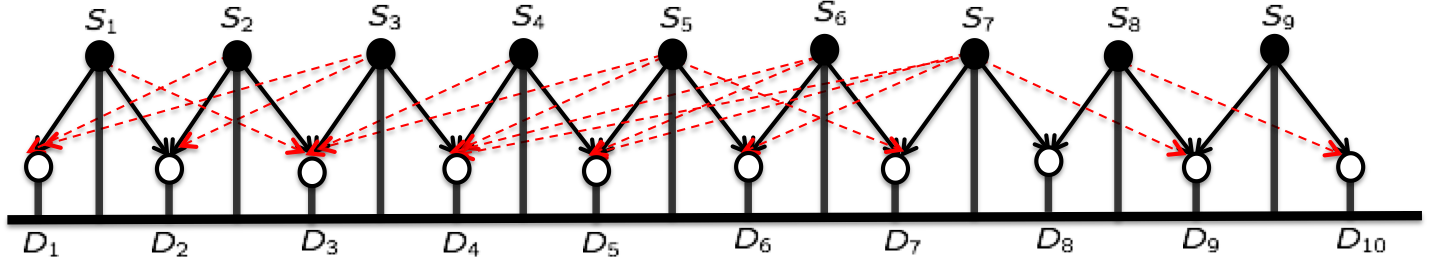}
\caption{\it A one-dimensional convex cellular network where the left-to-right greedy orthogonal scheme achieves 3 DoF by scheduling orthogonal transmissions $S_1\rightarrow D_1$, $S_4 \rightarrow D_4$ and $S_8\rightarrow D_8$. This is also the optimal DoF value.}
\label{fig:random1}
\end{figure}

\begin{figure}[!h] \centering
\includegraphics[width=6in]{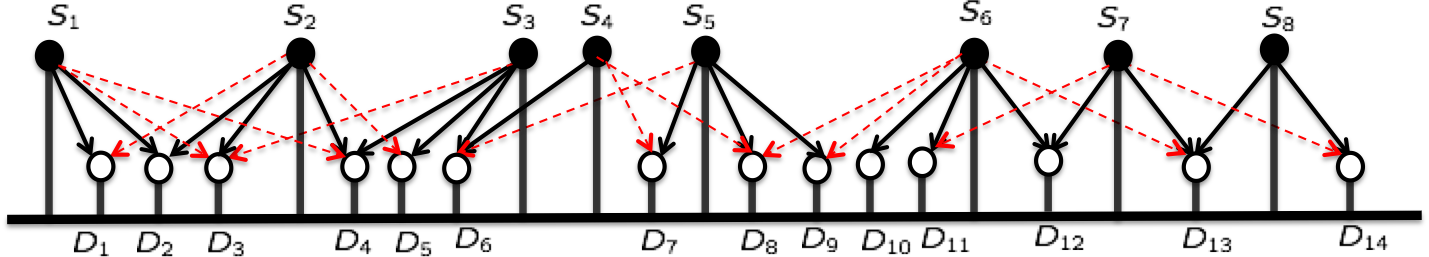}
\caption{\it A one-dimensional convex cellular network with convex connectivity where the left-to-right greedy orthogonal scheme achieves 5 DoF by scheduling orthogonal transmissions $S_1\rightarrow D_1$, $S_3\rightarrow D_5$, $S_5\rightarrow D_7$, $S_6\rightarrow D_{10}$, and $S_8\rightarrow D_{14}$. This is also the optimal DoF value.}
\label{fig:random2}
\end{figure}

Two examples of one-dimensional convex cellular networks are illustrated in Fig. \ref{fig:random1} and Fig. \ref{fig:random2} where the source nodes $S_i$ are the tall black towers and the destination nodes $D_j$ are short white towers. Solid black edges show cell associations, i.e., there is a solid black edge from a source node to a destination node if and only if there is a desired message between the two. Dashed red edges show interference, i.e., a source is connected to a destination by a dashed edge if the channel between them is strong enough for the source to be heard by the destination but there is no desired message between them. If there is no edge from a source to a destination then it means that the destination cannot hear the source, modeled as a zero channel between them. It is easy to check that these are one-dimensional convex cellular networks. For such networks, we will define a greedy orthogonal access scheme.

\subsection{Left-to-Right Greedy Orthogonal Access Scheme}
As the name suggests, the greedy orthogonal access scheme  will greedily select an orthogonal subset of messages for transmission. 

\bigskip

\noindent{\bf Orthogonal messages:} These are messages that cause no interference to each other, i.e., a subset of messages is called orthogonal if out of all the sources and destinations that are associated with these messages, the intended destination of any message can hear only one source, which transmits only its desired message. 

\bigskip
Clearly, an orthogonal set of messages cannot include more than one message intended for the same destination or originating at the same source.
%\begin{definition}[Orthogonal Messages] A subset of messages $\mathcal{W}_o\subset\mathcal{W}$ is considered orthogonal if and only if
%\begin{eqnarray}
%\mathcal{D}(W_j)\nrightarrow\mathcal{S}(W_i), \forall W_i, W_j\in\mathcal{W}_o, i\neq j
%\end{eqnarray}
%\end{definition}

The left-to-right greedy scheme starts at the left and moves to the right, building a set of orthogonal messages by greedily adding any message that is orthogonal to its previously chosen set of orthogonal messages.  It starts with the first source and first destination, i.e.,  source $S_1$, and destination $D_{1}$. Because of the network convexity, and since every destination must have at least one desired message, $S_1$ must have a message for $D_1$. This message is the first message chosen by the greedy orthogonal scheme. The remaining messages are chosen by moving to the right in the following iterative fashion. Suppose at any stage, the last chosen orthogonal message was from $S_i$ to $D_j$. Then the greedy scheme looks for the first source to the right of $S_i$ that cannot be heard by $D_j$. When it finds such a source, say $S_k$, then it looks among the desired destinations of $S_k$ to find the first destination that does not hear $S_i$.  If it finds such a destination, say $D_l$, then it adds the message from $S_k$ to $D_l$ as the next member of the orthogonal set. If it does not find such a destination, then it moves on to the next source on the right, $S_{k+1}$, searches for the first among its desired destinations that does not hear $S_i$, and so on. If it finds such a source-destination pair, the corresponding message is added to the orthogonal set and the search continues for the next message. The search stops when there are no more source nodes on the right. 

Let us illustrate how the greedy orthogonal scheme functions, through the two examples shown in Fig. \ref{fig:random1} and Fig. \ref{fig:random2}. First consider Fig. \ref{fig:random1}. After choosing $S_1\rightarrow D_1$, the next source to the right that cannot be heard by $D_1$ is $S_4$. The first desired destination of $S_4$ that does not hear $S_1$ is $D_4$. So the next orthogonal message chosen corresponds to $S_4\rightarrow D_4$. Continuing to the right, the next choice is $S_8\rightarrow D_8$ and there are no further choices available. Since three orthogonal messages are chosen for transmission, the DoF achieved by the left-to-right greedy orthogonal scheme is 3. Now consider Fig. \ref{fig:random2}. The left-to-right greedy orthogonal scheme for this network chooses the orthogonal messages corresponding to $S_1\rightarrow D_1$, $S_3\rightarrow D_5$, $S_5\rightarrow D_7$, $S_6\rightarrow D_{10}$, and $S_8\rightarrow D_{14}$ to achieve a total of 5 DoF.

\subsection{Optimality of Left-to-Right Greedy Orthogonal Access Scheme}
One might wonder if we could have done better with an orthogonal scheme that moved from  right-to-left, instead of left-to-right. Indeed the question is non-trivial. We will answer this question rigorously very shortly, but  perhaps it is interesting to first try it out. For the network of Fig. \ref{fig:random1}, a right-to-left greedy orthogonal scheme will choose $S_9\rightarrow D_{10}$, $S_7\rightarrow D_8$ and $S_3\rightarrow D_3$, to again achieve a total of $3$ DoF. Similarly, for the network of Fig. \ref{fig:random2}, a right-to-left greedy orthogonal scheme will choose $S_8\rightarrow D_{14}$, $S_6\rightarrow D_{12}$, $S_5\rightarrow D_7$, $S_3\rightarrow D_5$, and $S_1\rightarrow D_2$, to again achieve a total of 5 DoF. One might also wonder if non-greedy schemes may be able to find a larger set of orthogonal messages, and it may also be an interesting exercise. However, as we will show next, no  scheme can find a larger set of orthogonal messages than the left-to-right greedy orthogonal scheme. In fact, we will show a \emph{much} stronger result. That no scheme, orthogonal or otherwise, linear or non-linear, can achiever greater DoF than the left-to-right greedy orthogonal scheme, i.e., the left-to-right greedy orthogonal scheme achieves the information theoretic optimal DoF. This is the main result of this paper, and is stated in the following theorem.
\begin{theorem}
\label{theorem1}
The left-to-right greedy orthogonal access scheme achieves the optimal information theoretic sum DoF of all one-dimensional convex cellular networks.
\end{theorem}
{\it Remark:} Note that Theorem \ref{theorem1} applies only to the topological interference management setting, i.e., no CSIT is assumed to be available other than the knowledge of the network topology. Also, since the choice of left-to-right direction is arbitrary, the corresponding right-to-left greedy orthogonal scheme must be DoF-optimal as well.

While the cellular networks considered in this work are illustrated as downlink scenarios (base stations transmitting), all uplink settings (base stations receiving) are automatically included by switching the notion of users and base stations, i.e., if we regard the sources as users and destinations as base stations, then all the networks become uplink networks. Since the labeling of sources and destinations as base stations or users is entirely a cosmetic issue, the result of Theorem \ref{theorem1} applies to both uplink and downlink settings. 

Note that for any convex cellular network, the reciprocal network (obtained by switching the direction of communication) is also a convex cellular network. Therefore, the greedy orthogonal access scheme is sum-DoF optimal  in both networks. The greedy orthogonal access scheme  achieves  the same sum-DoF in both networks, so from its optimality it follows that every one-dimensional convex-cellular network has the same sum-DoF as its reciprocal network, i.e., duality holds. 

Last but not the least, the result of Theorem \ref{theorem1} extends to corresponding instances of the index coding problem, leading to the following corollary.

\begin{corollary}
\label{corollary1}
The left-to-right greedy orthogonal access scheme achieves the sum capacity of the index coding instances corresponding to one-dimensional convex cellular networks (as identified in \cite{Jafar_TIM}).
\end{corollary}

The proofs for Theorem \ref{theorem1} and Corollary \ref{corollary1} are presented in Section \ref{sec:proofs}.
\section{Discussion}
It is remarkable that orthogonal access schemes are optimal for all one-dimensional convex cellular networks because this is not the case for  two-dimensional or non-convex networks. 

One-dimensional network topology by itself does not imply that orthogonal access schemes are DoF optimal. For example, the 4 cell network of Fig. \ref{fig:4cell} can be  mapped to a one-dimensional topology by numbering the nodes and placing them sequentially, without changing the connectivity or message sets. But orthogonal access schemes are not optimal there. 

Convex cellular model  by itself also does not imply that orthogonal access  schemes are DoF optimal. For example, the 4 cell setting of Fig. \ref{fig:4cell} obviously is a convex cellular model in 2-dimensions, but orthogonal solutions are not optimal there. 

Note that the 4-cell example of Fig. \ref{fig:4cell} can be seen as a one-dimensional non-convex cellular network or as a two-dimensional convex cellular network, but not as a one-dimensional convex cellular network.

Thus, taken individually, neither one-dimensional topology nor convex connectivity implies the optimality of orthogonal access schemes. It is therefore somewhat surprising that when considered together, the DoF of one-dimensional convex cellular networks are always achieved by orthogonal schemes. This observation is  the main contribution of this work. 

%Remarkably, we show that for this class of wireless (index coding) networks the DoF (capacity) optimal solution is an orthogonal scheme such as TDMA (clique cover). 

To understand the significance of this observation, note that one-dimensional topologies with convex connectivity, e.g., Wyner models \cite{Wyner}, are commonly used in information theoretic studies as canonical representatives of cellular networks \cite{Shamai_Wyner, Bergel_Yellin_Shamai, Somekh_Zaidel_Shamai, Sanderovich_Somekh_Poor_Shamai, Levy_cluster, Lapidoth_Levy_Shamai_Wigger, Gesbert_cluster,Xu_Zhang_Andrews} in order to gain fundamental insights into cellular interference management principles. While the limitations of such models have been explored from a practical perspective \cite{Xu_Zhang_Andrews}, what our result shows is a \emph{fundamental} limitation of these canonical models. Specifically, the study of one dimensional convex  topologies in the absence of CSIT, cannot reveal the need for non-orthogonal  access schemes, which become necessary for two-dimensional or non-convex topologies. Finally, the corresponding class of index coding problems whose capacity is automatically characterized, is interesting as well.

\section{System Model}
\subsection{Channels and Message Sets}
Consider a one-dimensional downlink linear cellular network where the source nodes (transmitters) are denoted from left to right as $S_1, S_2,$ $\cdots, S_T$  and there are totally $K$ users (destinations) that are sequentially denoted, from left to right, as $D_1,D_2,\ldots,D_K$, respectively. 
The channel input-output relationships are denoted as
\begin{eqnarray}
\left[\begin{array}{c}
y_1(n)\\
y_2(n)\\
\vdots \\
y_K(n)\end{array}\right]=
\left[\begin{array}{cccc}
h_{11}(n)&h_{12}(n)&\ldots&h_{1T}(n)\\
h_{21}(n)&h_{22}(n)&\ldots&h_{2T}(n)\\
\vdots&\vdots&\ddots&\vdots\\
h_{K1}(n)&h_{K2}(n)&\ldots&h_{KT}(n)
\end{array}\right]
\left[\begin{array}{c}
x_1(n)\\
x_2(n)\\
\vdots\\
x_T(n)
\end{array}\right]+
\left[\begin{array}{c}
z_1(n)\\
z_2(n)\\
\vdots\\
z_K(n)
\end{array}\right]
\end{eqnarray}
where, over the $n$-th channel use, $x_j(n)$ is the transmitted symbol from source (base station) $S_j$ , $y_i(n)$ is the received symbol at User $D_i$, $z_i(n)\sim\mathcal{N}^c(0,1)$ is zero mean unit variance additive complex Gaussian noise at destination (user) $D_i$, and $h_{ij}(n)$ is the channel coefficient between source $S_j$ and destination $D_i$. Under the TIM problem the CSIT is a binary value for each channel that models it as either connected or disconnected. The disconnected channel coefficients are zero. The connected channels  take values in a range bounded away from zero and infinity but their values are not known to the transmitters. The non-zero (connected) channel coefficients  may be assumed to remain constant or vary across time, but are statistically indistinguishable from each other, as far as the transmitters are concerned. Channel state information at the receivers includes, in addition to the topological information, the precise knowledge of all desired channels. All symbols belong to the field of complex numbers $\mathbb{C}$.

We assume a  general model where each user may receive desired messages from possibly multiple but not necessarily all the base stations that it can hear. This includes as a special case the conventional setting where each user is only associated with one desired base station, e.g., the closest base station, and receives only interference from the remaining base stations that it can hear, but is general enough to  also model soft-handoff and other cooperative scenarios where multiple base stations send information to each user. Thus, base station $S_j$ has a set of independent messages, $\mathcal{W}(S_j)$, that it wants to send to its desired users. User $D_i$ has a set of independent messages $\mathcal{W}(D_i)$ that it desires. The set of all independent messages is denoted as $\mathcal{W}$, so that
\begin{eqnarray}
\mathcal{W}=\cup_{i=1}^{K}\mathcal{W}(D_i)=\cup_{i=1}^{T}\mathcal{W}(S_i).
\end{eqnarray}
Each  message $W\in\mathcal{W}$ has a unique source $\mathcal{S}(W)$ and a unique destination $\mathcal{D}(W)$. In other words, this is a multiple unicast setting.

Between any base station $S_i$ and user $D_j$, we have three possibilities. 
\begin{enumerate}
\item {[}{\bf Weak}  ($D_j \nrightarrow S_i$){]} \\The channel can be weak (zero). This is denoted as $D_j \nrightarrow S_i$. Note the direction points from the destination to the source. This is motivated by the notion of acyclic demand graphs \cite{Jafar_TIM} that will be useful later on. In the illustrations, we will  indicate weak channels  simply by the lack of an edge between the base station and the user. 
%However, occasionally when we need to highlight a specific weak channel, we will indicate it with an edge with a strikethrough. 
\item {[}{\bf Desired} ($S_i\rightarrow D_j$) {]} \\If the channel is non-zero, and $S_i$ has a desired message for $D_j$ then we call it a  desired channel and denote it as $S_i\rightarrow D_j$.  In the figures, such a channel is shown as a solid black edge between $S_j$ and $D_i$.  
\item {[}{\bf Interfering} ($S_i\dashrightarrow D_j$){]} \\ If $S_i$ can be heard by $D_j$ (non-zero channel) but there is no desired message from $S_i$ to $D_j$, then we call it an interfering channel and denote it as $S_i\dashrightarrow D_j$. In the figures, such a channel is  indicated by a dashed red edge.
\end{enumerate}
To avoid degenerate cases, we will assume without loss of generality that if a source has a message for a destination, the channel between them cannot be zero.
\begin{eqnarray}
\mbox{If } S_i\rightarrow D_j, \mbox{ then it cannot be true that } D_j\nrightarrow S_i. \label{eq:degenerate}
\end{eqnarray}

%
%
%We will assume the following throughout this work.
%\begin{itemize}
%\item The channel coefficient values are assumed to can be fixed or i.i.d during the communication.
%\item The topology of the network is known to all base stations and users
%\item Besides the topology information, there is no CSIT.
%\item Besides the topology information, the CSIR only includes the knowledge of the desired channel coefficient at each receiver.
%\end{itemize}
%This assumption is valid because the connectivity always changes much slower than the channel gains. 
%All symbols are complex, and the $z_i(n)$ terms represent additive white Gaussian noise, independent identically distributed $~N^c(0,N_0)$.

 \noindent The  transmit power constraint at every base station  is set as $P$ , i.e.,
\begin{align}
\text{E} \left[\frac{1}{N}\sum_{n=1}^{N}|x_i(n)|^2\right] \leq P, ~~~\forall i\in\{1,2,\cdots, T\}.
\end{align}

\noindent To compute the DoF, we let $P$ approach infinity, and evaluate the achievable rates normalized by $\log(\text{P})$. If there exists a sequence of achievable rate allocations $R(W, P)$, such that the limit $R(W,P)/ \log(P)$ exists for all $W \in \mathcal{W}$ as $P \rightarrow \infty$, then these limiting values are said to be an achievable DoF allocation.
\begin{eqnarray}
\text{DoF}(W)=\lim_{P\rightarrow \infty} \frac{R(W,P)}{\log(P)}~ \forall W \in \mathcal{W}
\end{eqnarray}
The closure of achievable DoF allocations is called the DoF region. Our goal is to find the sum-DoF in such a network, i.e.,  the maximum value of the sum of DoF of all the messages in the network. 

%Hereafter, when we say a base station is connected to a user, it means that it is connected through a channel that is above the noise floor.

%, to ensure the following nominal interference-free SNR guarantees for all desired links:
%\begin{eqnarray}
%\label{channel}
%\frac{|h_{ij}|^2P_j}{N_0}\geq \text{SNR}~\forall i \in \{1,2,\ldots,K\}, j \in \{1,2,\ldots,T\},~ \mathcal{W}_{D_i}\cap \mathcal{W}_{S_j}\neq \emptyset
%\end{eqnarray}
%
%Thus, the power constraints are chosen such that, in the absence of all other messages, each message by itself can achieve a rate $\log(1 + \text{SNR})$. Since (\ref{channel}) is the only information available to the
%transmitters about desired channels, $\log(1 + \text{SNR})$ is also the individual capacity of each message if all other messages are eliminated, i.e., allocated zero rates.
%
%In addition to the interference-free SNR guarantees for the desired channels, as in (\ref{channel}), let us allow just one bit of CSIT about the interference channel strengths. A natural choice for this one bit CSIT could be as follows. The receivers compare the nominal received power from the undesired links to a pre-chosen threshold value, which is effectively the acceptable noise floor, and assign a $'0'$ to all those (weak) links whose collective contribution is below the noise floor. All other significant links are assigned the value $'1'$. These assignments comprise the 1-bit CSIT of the undesired channel strengths.

\subsection{Convex Cellular Model}
We will specialize the convex cellular model to the one-dimensional case. But  first we need the notion of relative node position, defined as follows.
\begin{definition} We define the relation $a < b$ between a source node and a destination node to indicate that node $a$ is ``to the left of" node $b$. For example, $S_1<D_1$ would mean that source 1 is to the left of destination 1, $D_2<S_2$ would mean that destination 2 is to the left of source 2. Among source nodes $S_i<S_j$ if $i<j$ and similarly among destination nodes $D_i<D_j$ if $i<j$. The notations $a\leq b$, $a >b$, $a\geq b$, when applied directly to source and/or destination nodes,  are defined similarly.
\end{definition}
The one-dimensional convex cellular model is defined by destination and source convexity properties listed below.
\begin{enumerate}
\item {\bf Destination Convexity:} Destination convexity refers to the property that if a destination has a desired message from a source node on its left (right) side, then it must also have a desired message from all other source nodes on the left (right) side that are closer, and if a destination cannot hear a source node on its left (right) side, then it must also be unable to hear all other sources on the left  (right) side that are farther away. This is expressed notationally as follows.
\begin{figure}[!h] \centering
\includegraphics[width=5in]{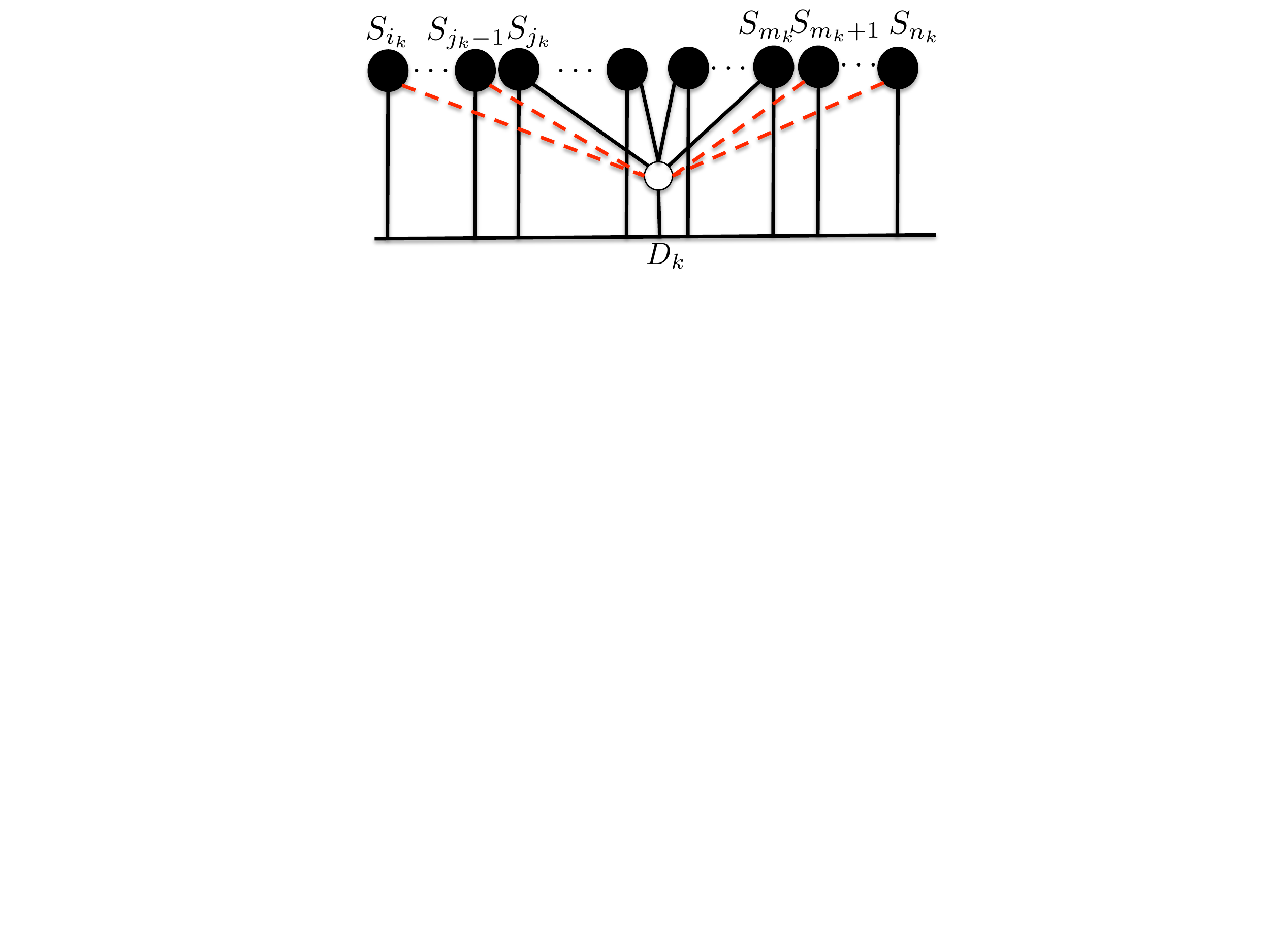}
\vspace{-200pt}
\caption{\it Destination $D_k$ has independent desired messages from sources connected with black links and receives only  interference from sources connected with red links.}
\label{fig:pathloss}
\end{figure}
\begin{enumerate}
\item If $S_i<S_j<D_k$, and $S_i\rightarrow D_k$ then it is implied that $S_j\rightarrow D_k$. 
\item If $S_i>S_j>D_k$, and $S_i\rightarrow D_k$ then it is implied that $S_j\rightarrow D_k$. 
\item If $S_i<S_j<D_k$, and $D_k\nrightarrow S_j$ then it is implied that $D_k\nrightarrow S_i$. 
\item If $S_i>S_j>D_k$, and $D_k\nrightarrow S_j$ then it is implied that $D_k\nrightarrow S_i$. 
\end{enumerate}
Destination convexity is shown in Fig. \ref{fig:pathloss} from the perspective of destination $D_k$.  Base stations $S_{j_k}$ through $S_{m_k}$ have independent messages for user $D_{k}$. The base stations that are out of this range but still close enough to send  interference are shown as two subsets of base stations in Fig. \ref{fig:pathloss}. One of the subsets is comprised of base stations $S_{m_k+1}$ through $S_{n_k}$ and the other subset is comprised of base stations $S_{i_k}$ through $S_{j_k-1}$ that are respectively located on the right and left side of the desired set of base stations. The remaining base stations (not shown in Fig. \ref{fig:pathloss}) present only weak (zero) channels to user $D_k$. Note that no symmetry is assumed in the ranges of desired, interfering and weak base stations.

\item {\bf Source Convexity:}
Source convexity refers to the property that if a source has a desired message for a destination node on its left (right) side, then it must also have a desired message to all other destination nodes on the left (right) side that are closer, and if a source cannot be heard by a destination node on its left (right) side, then it must also be unable to be heard by all other destination nodes on the left (right) side that are farther away. This is expressed notationally as follows.
%\begin{figure}[!h] \centering
%\includegraphics[width=4in]{pathloss1}
%\vspace{-80pt}
%\caption{Destination $D_k$ has independent messages from sources connected with black links and get interfered with base stations connected with red links.}
%\label{fig:pathloss}
%\end{figure}
\begin{enumerate}
\item If $D_i<D_j<S_k$, and $S_k\rightarrow D_i$ then it is implied that $S_k\rightarrow D_j$. 
\item If $D_i>D_j>S_k$, and $S_k\rightarrow D_i$ then it is implied that $S_k\rightarrow D_j$. 
\item If $D_i<D_j<S_k$, and $D_j\nrightarrow S_k$ then it is implied that $D_i\nrightarrow S_k$. 
\item If $D_i>D_j>S_k$, and $D_j\nrightarrow S_k$ then it is implied that $D_i\nrightarrow S_k$. 
\end{enumerate}

\begin{figure}[!h] \centering
\includegraphics[width=5in]{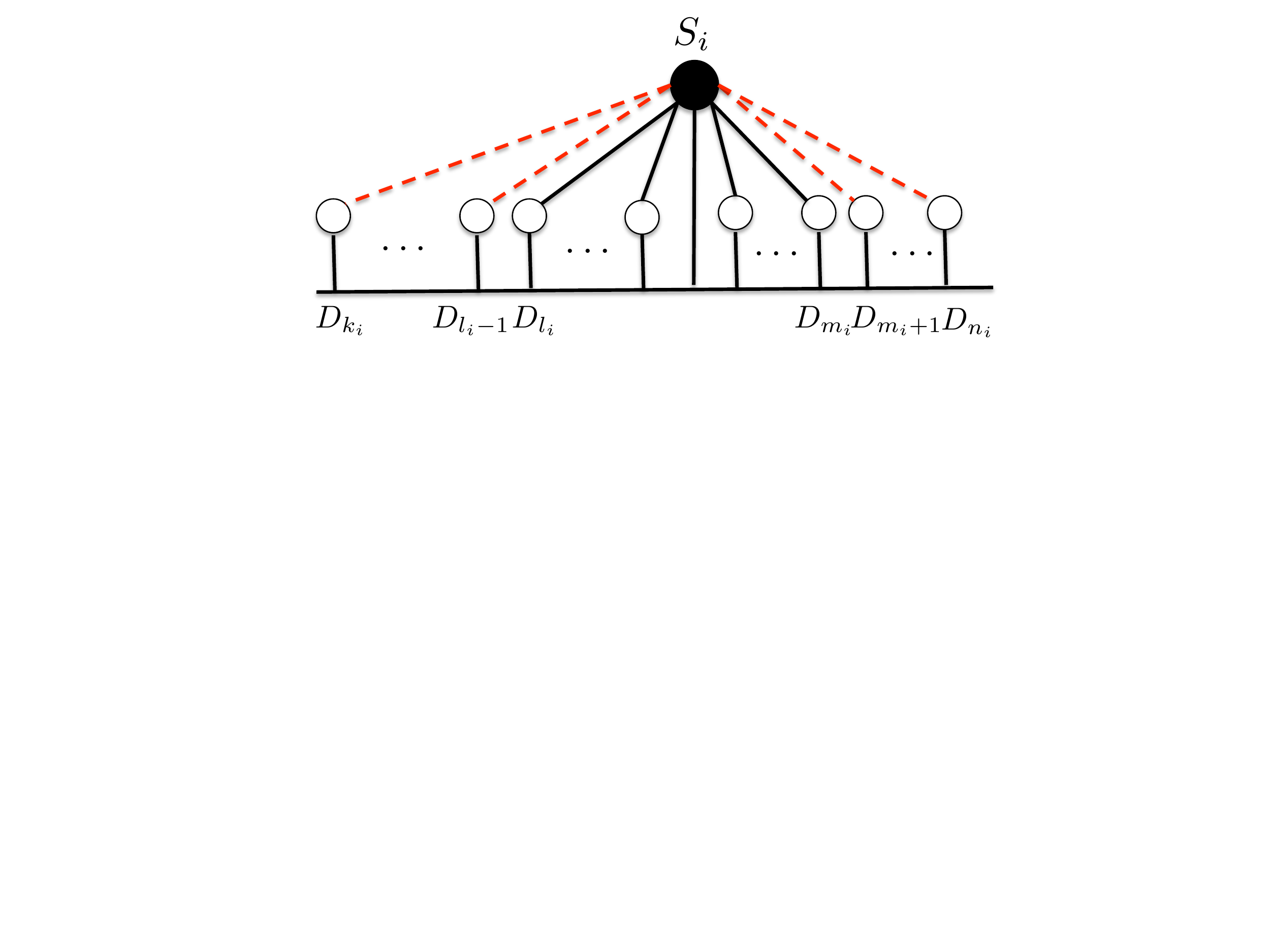}
\vspace{-180pt}
\caption{Source $S_i$ has independent messages for the destinations connected with black links and interferes with users connected with red links.}
\label{fig:pathloss2}
\end{figure}
Source convexity is shown in Fig. \ref{fig:pathloss2} from the perspective of source $S_i$.  Destinations $D_{l_i}$ through $D_{m_i}$ receive independent messages from source  $S_{i}$. The destinations that are out of this range but still close enough to receive  interference are shown as two subsets of destinations in Fig. \ref{fig:pathloss2}. One of the subsets is comprised of destinations $D_{m_i+1}$ through $D_{n_i}$ and the other subset is comprised of destinations $D_{k_i}$ through $D_{l_i-1}$ that are respectively located on the right and left side of the desired set of destinations. The remaining destinations (not shown in Fig. \ref{fig:pathloss2}) see only weak (zero) channels from source $S_i$. Once again, note that no symmetry is assumed in the ranges of desired, interfering and weak base stations.

\end{enumerate}
{\it Remark:} It is easy to verify that destination convexity does not imply source convexity and vice versa, i.e., the two are not redundant.

\section{Proof}\label{sec:proofs}
\subsection{Preliminaries}
\begin{lemma} \label{lemma:cycle}
Let $\mathcal{W}_a\subset \mathcal{W}$ be a subset of messages,  $\mathcal{S}(\mathcal{W}_a)$ be the set of source  nodes  where these messages originate and $\mathcal{D}(\mathcal{W}_a)$ be the set of destination nodes for which these messages are intended. If the sum-DoF value for the messages in $\mathcal{W}_a$ is greater than 1, then there must exist a cycle of the form:
\begin{eqnarray}
D_{i_1}\nrightarrow S_{i_2}\rightarrow D_{i_3}\nrightarrow S_{i_4}\rightarrow\cdots\nrightarrow S_{i_n}\rightarrow D_{i_1}\label{eq:cycle}
\end{eqnarray}
with $S_{i_2}, S_{i_4}, \cdots, S_{i_n} \in \mathcal{S}(\mathcal{W}_a)$ and $D_{i_1}, D_{i_3},\cdots, D_{i_{n-1}}\in\mathcal{D}(\mathcal{W}_a)$.
\end{lemma}
\proof Proof follows directly from Theorem 4.11 of \cite{Jafar_TIM} which states that the sum-DoF value must be 1 if the demand-graph of a TIM problem is acyclic. Therefore, if the sum-DoF value is greater than 1, the demand graph must contain a cycle. This is the cycle indicated in (\ref{eq:cycle}).\hfill\QED

\begin{lemma}\label{lemma:wrap} Let $\mathcal{W}_a\subset \mathcal{W}$ be a subset of messages,  $\mathcal{S}(\mathcal{W}_a)$ be the set of source  nodes  where these messages originate and $\mathcal{D}(\mathcal{W}_a)$ be the set of destination nodes for which these messages are intended. If the sum-DoF value for the messages in $\mathcal{W}_a$ is greater than 1, then there must exist 
\begin{eqnarray}
D_{j_1}\nrightarrow S_{j_2}\rightarrow D_{j_3}\nrightarrow S_{j_4}\rightarrow D_{j_5}\label{eq:wrap}
\end{eqnarray}
with $S_{j_2}, S_{j_4} \in \mathcal{S}(\mathcal{W}_a)$ and $D_{j_1}, D_{j_3}, D_{j_{5}}\in\mathcal{D}(\mathcal{W}_a)$ such that 
\begin{eqnarray}
D_{j_1}&<&D_{j_3}\label{eq:hopright}\\
D_{j_5}&<&D_{j_3}\label{eq:hopleft}\\
S_{j_4}&<&D_{j_3}\label{eq:sourceleft}\\
D_{j_1}&<&S_{j_2}\label{eq:sourceright}
\end{eqnarray}
\end{lemma}
\proof According to Lemma \ref{lemma:cycle} a cycle such as (\ref{eq:cycle}) must exist. Let us start with the left-most destination node in that cycle and consider only each subsequent destination node that we encounter (ignore source nodes for now).  Since we started with the left-most destination node we must initially move to the right in terms of destination nodes as we traverse the cycle. However, since this is a cycle and must eventually return back to the left-most destination node which was our starting point, there must be a  destination node (call it $D_{j_3}$) such that the previous destination node that we encountered before $D_{j_3}$  (call it $D_{j_1}$) was to the left of $D_{j_3}$ and also the next destination node that we encounter after $D_{j_3}$ (call it $D_{j_5}$) is to the left of $D_{j_3}$, making a U-turn at $D_{j_3}$. For this choice of $D_{j_1}, D_{j_3}, D_{j_5}$, (\ref{eq:hopright}) and (\ref{eq:hopleft}) must hold. Note that it does not matter if $D_{j_1}$ and $D_{j_5}$ are the same node, which is a possibility. 

Now in order to prove (\ref{eq:sourceleft}) let us assume, by way of contradiction, that $D_{j_3}<S_{j_4}$. Because of (\ref{eq:hopleft}) we must have $D_{j_5}<D_{j_3}<S_{j_4}$. Then using  (\ref{eq:wrap})  and by  property (c) of source convexity, since source $S_{j_4}$ cannot be heard by $D_{j_3}$, i.e., $D_{j_3}\nrightarrow S_{j_4}$, and $D_{j_5}$ is even farther away, $D_{j_5}$ must also be unable to hear source $S_{j_4}$, i.e., $D_{j_5}\nrightarrow S_{j_4}$, but this contradicts (\ref{eq:wrap}) because according to (\ref{eq:wrap}), source $S_{j_4}$ must have a desired message for destination $D_{j_5}$, i.e., we must have $S_{j_4}\rightarrow D_{j_5}$. Recall that according to (\ref{eq:degenerate}) if a source has a message for a destination, then the channel between them must be non-zero. The contradiction implies that the assumption that $D_{j_3}<S_{j_4}$ cannot be correct, thus proving (\ref{eq:sourceleft}).

Finally in order to prove (\ref{eq:sourceright}) let us assume, again by way of contradiction, that $D_{j_1}>S_{j_2}$. Because of (\ref{eq:hopright}) we must have $D_{j_3}>D_{j_1}>S_{j_2}$. But since according to (\ref{eq:wrap}), $D_{j_1}\nrightarrow S_{j_2}$, i.e., source $S_{j_2}$ is not heard by destination $D_{j_1}$ on its right, and $D_{j_3}$ is even further away on the right, by property (d) of source convexity, $D_{j_3}$ must not be able to hear $S_{j_2}$ either, i.e., $D_{j_3}\nrightarrow S_{j_2}$. But this contradicts (\ref{eq:wrap}) which requires $S_{j_2}\rightarrow D_{j_3}$. The contradiction implies that the assumption that $D_{j_1}>S_{j_2}$ cannot be correct, thus proving (\ref{eq:sourceright}).\hfill\QED

\begin{lemma}\label{lemma:start}
For a one-dimensional convex connected network, where after $S_1\rightarrow D_1$, the left-to-right greedy orthogonal scheme would choose $S_i\rightarrow D_j$, consider the set of all messages that either originate from $S_1, S_2, \cdots, S_{i-1}$, or are intended for destinations $D_1, D_2, \cdots, D_{j-1}$. Then, the information-theoretic sum-DoF of all these messages cannot be more than 1.
\end{lemma}
\proof The proof is by contradiction. Suppose the sum-DoF of these messages is  greater than 1. Then, by Lemma \ref{lemma:wrap}, there must exist destinations $D_{j_1}, D_{j_3}, D_{j_5}$ that are either among $D_1, D_2, \cdots, D_{j-1}$ or are the intended recipients of messages originating at sources $S_1, S_2, \cdots, S_{i-1}$, and  there must exist sources $S_{j_2}, S_{j_4}$ that are either among the sources $S_1, S_2, \cdots, S_{i-1}$ or are the origins of messages that are intended for a destination that is among $D_1, D_2, \cdots, D_{j-1}$, such that (\ref{eq:wrap})-(\ref{eq:sourceright}) are satisfied.
%\begin{eqnarray}
%D_{j_1}\nrightarrow S_{j_2}\rightarrow D_{j_3}\nrightarrow S_{j_4}\rightarrow D_{j_5}\label{eq:wrap}
%\end{eqnarray}
%and 
%\begin{eqnarray}
%D_{j_1}&<&D_{j_3}\label{eq:hopright}\\
%D_{j_5}&<&D_{j_3}\label{eq:hopleft}\\
%S_{j_4}&<&D_{j_3}\label{eq:sourceleft}\\
%D_{j_1}&<&S_{j_2}\label{eq:sourceright}
%\end{eqnarray}

Now, because $D_{j_1}\nrightarrow S_{j_2}$ and $D_1\leq D_{j_1}<S_{j_2}$, because of the convexity of sources we must have
\begin{eqnarray}
D_1\nrightarrow S_{j_2}\label{eq:choose1}
\end{eqnarray}
and because $D_{j_3}\nrightarrow S_{j_4}$ and $S_1\leq S_{j_4}<D_{j_3}$,  because of the convexity of destinations we must have
\begin{eqnarray}
D_{j_3}\nrightarrow S_{1}\label{eq:choose2}
\end{eqnarray}
According to (\ref{eq:choose1}) and (\ref{eq:choose2}) the message from $S_{j_2}\rightarrow D_{j_3}$ is orthogonal to the message from $S_1\rightarrow D_1$. 

Suppose $S_{j_2}\in\{S_1, S_2,\cdots, S_{i-1}\}$. Then, because $S_{j_2}<S_i$, we have a contradiction because the left-to-right greedy orthogonal scheme should have picked $S_{j_2}\rightarrow D_{j_3}$ instead of $S_i\rightarrow D_j$ as the next orthogonal message after $S_1\rightarrow D_1$. 

So we must have $S_{j_2}\notin \{S_1, S_2, \cdots, S_{i-1}\}$, i.e., $S_{j_2}\geq S_{i}$. This implies that  $D_{j_3}\in \{D_1, D_2, \cdots, D_{j-1}\}$.

Because the left-to-right greedy orthogonal scheme chose $S_i\rightarrow D_j$,  and we know that $D_{j_3}<D_j$ and  $D_{j_3}\nrightarrow S_1$, it follows that 
\begin{eqnarray}
S_i\mbox{ must not have a desired message for }D_{j_3}.\label{eq:contradict1}
\end{eqnarray}
Since $S_i$ has a message for $D_j$ but not for $D_{j_3}$, and $D_{j_3}<D_j$, it follows by the convexity of sources that $S_i > D_{j_3}$. 

But now we have $S_{j_2}\geq S_i>D_{j_3}$, and $S_{j_2}\rightarrow D_{j_3}$, i.e., $S_{j_2}$ has a desired message for $D_{j_3}$. Therefore, by the convexity of destinations,
\begin{eqnarray}
S_i\mbox{ must  have a desired message for }D_{j_3}.\label{eq:contradict2}
\end{eqnarray}
Thus, (\ref{eq:contradict1}) and (\ref{eq:contradict2}) contradict each other, disproving our initial hypothesis that the sum-DoF of these messages is greater than 1. \hfill\QED

Finally we are in a position to prove Theorem \ref{theorem1}.
\subsection{Proof of Theorem \ref{theorem1}}
\proof Starting from the left, suppose the left-to-right greedy scheme chooses $S_{i_1}\rightarrow D_{j_1}$, $S_{i_2}\rightarrow D_{j_2}$, $S_{i_3}\rightarrow D_{j_3}, \cdots, S_{i_n}\rightarrow D_{j_n}$. 

Consider all the messages that either originate from $S_{i_1}, S_{i_1+1}, \cdots, S_{i_2-1}$ or are intended for destinations $D_{j_1}, D_{j_1+1}, \cdots, D_{j_2-1}$. By Lemma \ref{lemma:start} the information theoretic sum-DoF value of all these messages, through any achievable scheme, cannot be greater than 1.

Now, let us eliminate sources $S_{i_1}, S_{i_1+1}, \cdots, S_{i_2-1}$ and destinations $D_{j_1}, D_{j_1+1}, \cdots, D_{j_2-1}$ and all messages that either originate at or are intended for them. Clearly eliminating these cannot hurt the best achievable rates of the remaining messages. 

The new network starts with source $S_{i_2}$ as the left-most source node and destination $D_{j_2}$ as the left-most destination node. After $S_{i_2}\rightarrow D_{j_2}$ the left-to-right greedy orthogonal scheme would pick $S_{i_3}\rightarrow D_{j_3}$. Applying  Lemma \ref{lemma:start} again, we note that the sum-DoF of all messages that either originate from $S_{i_2}, S_{i_2+1}, \cdots, S_{i_3-1}$ or are intended for $D_{j_2}, D_{j_2+1}, \cdots, D_{j_3-1}$ cannot  exceed 1. 

Again eliminating these nodes and messages and repeating the argument $n$ times for each new network starting from $S_{i_1}\rightarrow D_{j_1}, \cdots, S_{i_n}\rightarrow D_{j_n}$, gives us a total of $n$ bounds on sum-DoF where every message has been accounted exactly once, and since each bound does not exceed 1, by adding them all we obtain the bound that the sum-DoF  of all the messages in the original network, achieved by \emph{any} scheme, cannot exceed $n$. However, $n$ is the sum-DoF value achieved by the left-to-right greedy orthogonal scheme. Therefore, the left-to-right greedy orthogonal scheme is DoF-optimal for all one-dimensional networks with convex connectivity.\hfill\QED

\subsection{Proof of Corollary \ref{corollary1}}
The outer bound of Theorem 1 is based only on the acyclic demand sets bound \cite{Jafar_TIM} which also applies to the corresponding index coding problem. The achievable scheme is based on orthogonal access which maps to a clique cover scheme in index coding. Thus the DoF result of Theorem \ref{theorem1} directly translates into a capacity result for the corresponding index coding problem.
\hfill\QED

\bibliography{Thesis}
\end{document}